# Valley Coherent Hot Carriers and Thermal Relaxation in Monolayer Transition Metal Dichalcogenides

*Sangeeth Kallatt[1,2,3], Govindarao Umesh[3], and Kausik Majumdar[1]\**

[1]Department of Electrical Communication Engineering, Indian Institute of Science, Bangalore 560012, India

[2]Center for NanoScience and Engineering, Indian Institute of Science, Bangalore 560012, India

[3]Department of Physics, National Institute of Technology Karnataka, Mangalore 575025, India

**Abstract:** We show room temperature valley coherence with in $MoS_2$, $MoSe_2$, $WS_2$ and $WSe_2$ monolayers using linear polarization resolved hot photoluminescence (PL), at energies close to the excitation – demonstrating preservation of valley coherence before sufficient scattering events. The features of the co-polarized hot luminescence allow us to extract the lower bound of the binding energy of the A exciton in monolayer $MoS_2$ as 0.42 ($\pm$0.02) eV. The broadening of the PL peak is found to be dominated by Boltzmann-type hot luminescence tail, and using the slope of the exponential decay, the carrier temperature is extracted *in-situ* at different stages of energy relaxation. The temperature of the emitted optical phonons during the relaxation process are probed by exploiting the corresponding broadening of the Raman peaks due to temperature induced anharmonic effects. The findings provide a physical picture of photo-generation of valley coherent hot carriers, and their subsequent energy relaxation path ways.

**Table of Contents (TOC) Graphic**

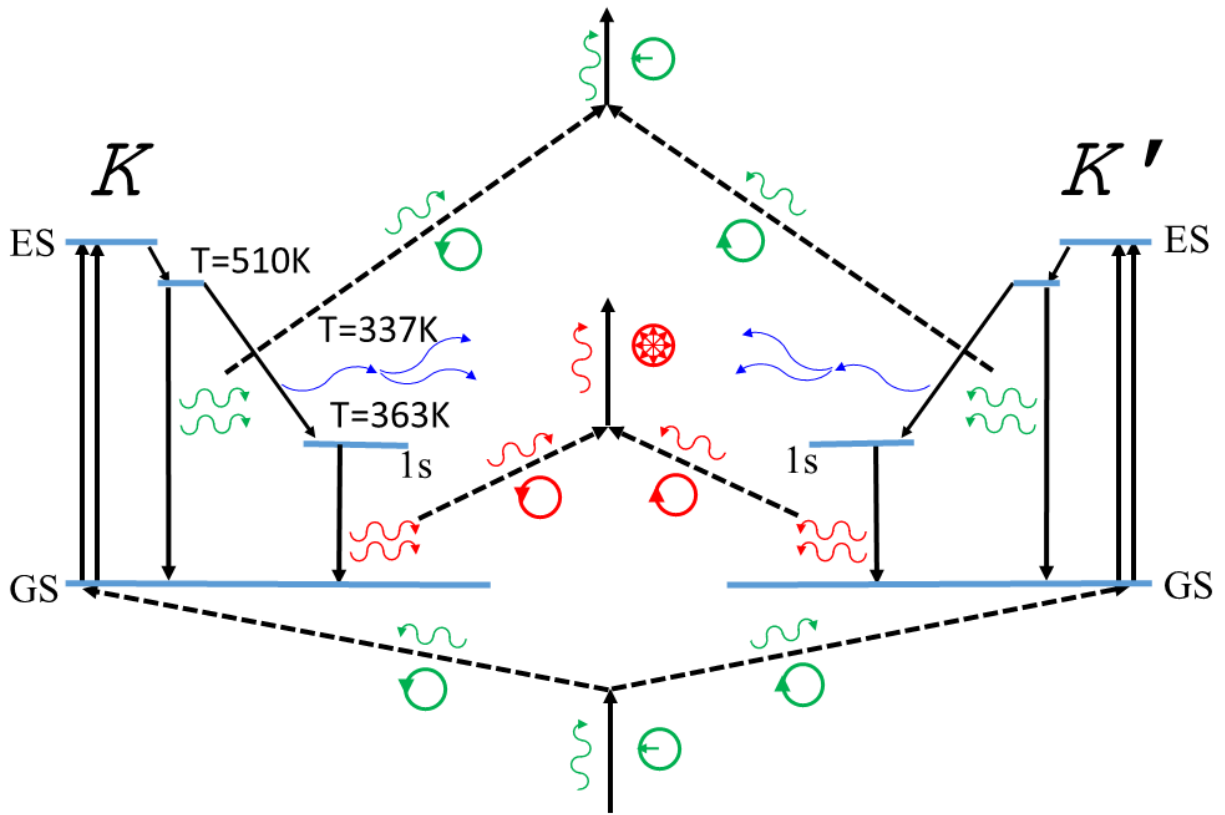

Monolayers of transition metal dichalcogenides (TMDs) exhibit excellent optical activity in spite of their sub-nm physical thickness ([1,2,3]), generating a lot of interest in 2D optoelectronic devices ([4,5,6,7]). Inversion symmetry broken monolayer TMDs show valley selective properties including circular dichroism ([8,9]), optical manipulation of valley hall effect ([10]), magnetic ([11,12]) and electrical ([13]) control of valley carriers. In addition, the exciton binding energy is extremely high (0.3-0.6 eV) ([14,15,16,17,18,19]) in these monolayers owing to their strong out of plane carrier confinement, large carrier effective mass and small dielectric constant. Recently, valley coherence has been reported with an excitation resonant with the excitonic bandgap ([20]). However, for off-resonant excitation, a physical understanding of the nature of valley coherence generated from parallel excitation of multiple excitonic levels and the subsequent energy relaxation is still lacking. Carriers generated through off-resonant excitation often play a key role in a variety of optoelectronic devices, and gaining insights into the nature of such photo-excited carriers and their eventual relaxation dynamics have important consequences from both fundamental science as well as technological points of view. Here we demonstrate room temperature valley coherence with off-resonant excitation in $MoS_2$, $MoSe_2$, $WSe_2$ and $WS_2$ monolayers using linear polarization resolved hot photoluminescence (PL), at energies close to the excitation. The results directly indicate preservation of valley coherence before sufficient optical phonon scattering. The features of the co-polarized hot luminescence allow us to extract the lower bound of the binding energy of the A exciton in monolayer $MoS_2$ as 0.42 ($\pm$0.02) eV. Using a combination of hot luminescence tail and anharmonic effects of Raman peaks, we also probe *in-situ* the carrier and the phonon temperatures at different stages of energy relaxation.

Monolayer TMDs are exfoliated on Si wafers covered by 285 nm SiO$_2$ (see Methods). The details of the monolayer material characterization for all the materials used in this work are provided in Supporting Information S1. The measurement setup is shown in Fig. 1(a) and all measurements are performed at 290 K. Fig. 1(b) shows the linear polarization resolved photoluminescence intensity from monolayer MoS$_2$ sample with vertically polarized excitation. As we observe from the figure, the A peak (at ~1.85 eV) and B peak (at ~2.0 eV), originating due to spin splitting in the valence band, do not exhibit any polarization resolved characteristics, owing to intrinsic decoherence of trions (*20*), and intra- and inter-valley scattering induced decoherence of excitons (*20,21,22,23*). On the contrary, strong polarization resolved PL is observed close to the excitation energy. In spite of the strong Raman signals from MoS$_2$ as well as Si substrate confounding the PL signal close to excitation energy, we are able to extract the PL contribution from the baseline envelop. Note that the different Raman peaks themselves display polarization resolved characteristics. For example, as shown in Fig. 1(c), the A$_{1g}$ peak is strongly polarized, while the E$^1_{2g}$ peak is not, which is in agreement with the symmetry of the corresponding vibrations (Supporting Information S2).

In Fig. 2(a), we show a magnified plot of the MoS$_2$ polarization resolved PL signal close to the excitation at 2.33eV, and the PL envelops are shown by dashed lines as guide to eyes. The co-polarized (VV) PL envelop and the corresponding degree of polarization is found to have a peak around ~2.29 eV, as shown in Fig. 2(a) and (d), respectively. The incident vertically polarized photon is a coherent superposition of left (LCP) and right (RCP) circularly polarized light, which in turn creates electron-hole (e-h) pairs in the K and K′ valleys. These e-h pairs recombine radiatively, either before or after exciton formation (*24*), providing back LCP and RCP luminescence at the respective valleys. Such LCP and RCP luminescence can generate linearly polarized light only when the whole process in the two valleys occur coherently. The observation of strong linear polarization resolved characteristics imply excellent coherence between the K and K' hot luminescence. This is expected close to the excitation energy due to lack of provision for sufficient scattering of the excited carriers. The degree of linear polarization drops significantly below 2.2eV, which is ~130meV (more than twice the optical phonon energy) below the excitation. Interestingly, such polarization resolved hot luminescence was observed in III-V semiconductors as well (25), although it should be noted that the requirement of valley coherence is absent in III-V semiconductor since we have only one ($\Gamma$) active valley.

In Fig. 2(f), we schematically show the different excitonic levels as a function of $\boldsymbol{k} = \boldsymbol{k}_e = -\boldsymbol{k}_h$ (hence, zero kinetic energy of the center of mass of the exciton) for parabolic bands [(26)]:

$$E_{ex}(\boldsymbol{k}) = E_g(\boldsymbol{k}) - R_y^* \frac{1}{\left(n_B - \frac{1}{2}\right)^2} \qquad (1)$$

with $E_g(\boldsymbol{k}) = E_{g0} + \frac{\hbar^2 k^2}{2m_e^*} + \frac{\hbar^2 k^2}{2m_h^*}$ being the quasiparticle bandgap at a given $\boldsymbol{k}$ without any excitonic effect and $\boldsymbol{k}$ is measured from the $\boldsymbol{K}$ ($\boldsymbol{K'}$) points. $E_{g0} = E_g(0)$ is the direct quasiparticle bandgap at the $\boldsymbol{K}$ ($\boldsymbol{K'}$) point. A Wannier exciton is assumed here as the calculated Bohr radius (~8Å) is larger than the unit cell dimension (~3.193Å). Each excitonic level $n_B$ can have a hot luminescence associated with a corresponding peak, when the excited electrons and holes of same $\boldsymbol{k}$ recombine back radiatively. Note that, lower the excitonic level, farther is the excitation point ($k_{n_B}$) from the K (K') point, as $k_{n_B} = \frac{1}{\hbar}\sqrt{2m^*(E_L - E_{g0} + \frac{R_y^*}{(n_B-\frac{1}{2})^2})}$ where $m^* = \frac{m_e^* m_h^*}{m_e^* + m_h^*}$ is the reduced mass of the exciton. For MoS$_2$, the calculated distance between the A$_{1s}$ excitation point and the K point is

~10% of the size of the Brillouin zone [Fig. 2(e)-(f)]. Using the prescription of ref. (*9*), the corresponding intrinsic depolarization associated with the excitation point being away from K (K′) point is calculated to be only 3%. Thus it seems more likely that the hot luminescence associated with each of the exciton branch is linearly polarized and contribute to the signal in Fig. 2(a). However, the observation of the strong peak around 2.29 eV followed by a decay is a signature of strong contribution from band extremum at K (K′). This is because the hot luminescence and its degree of polarization at $k \neq 0$ points should exhibit monotonically decaying characteristics from the excitation energy. The B exciton 2s and 3s levels have been reported to be around 2.24 and 2.34 eV (*17*), hence unlikely to contribute to the peak at 2.29 eV. Based on this, the 2.29 eV peak likely arises due to the A exciton higher energy level, which will lead to an estimated lower bound of the A exciton binding energy to be 0.42 ($\pm$0.02) eV. A summary of recently reported exciton binding energies of different monolayer TMDs is provided in supporting information S3.

The polarization resolved hot luminescence characteristics for monolayer $WSe_2$ and $WS_2$ are summarized in Fig. 2(b)-(e) and Fig. 2(g)-(h). For $WSe_2$, the monotonic decrement of the degree of polarization of the emitted light indicates absence of band extremum in the vicinity of the excitation energy – in agreement with A exciton bandgap of 1.65 eV and binding energy being ~0.3-0.4eV (*16*), as schematically shown in Fig. 2(g). On the other hand, in the case of $WS_2$, with A exciton bandgap around 2.05 eV and binding energy of 0.32 eV (*15,17*), the excitation is below the continuum level [Fig. 2(h)], and it exhibits increasing hot luminescence intensity with energy.

In Fig. 3, we have shown the difference between resonant and off-resonant excitation in $MoSe_2$ monolayer. When a 532 nm excitation is used [Fig. 3(a)-(c)], the situation is similar to $WSe_2$ case, and the 1s peak at ~1.57 eV remains unpolarized due to sufficient relaxation through scattering. Also, like other 2D materials, the observed photoluminescence is found to be linearly polarized close to the excitation energy. On the other hand, when the excitation wavelength is changed to 785 nm [Fig. 3(d)-(e)], we resonantly excite the MoSe$_2$ 1s level, which in turn exhibits very strong (>40%) linear polarization in the emergent photoluminescence – indicating strong valley coherent signal at room temperature. Only half of the 1s peak is observed in Fig. 3(e) due to the closeness of the edge filter associated with the resonant excitation.

Above the continuum limit band edge, the PL intensity corresponds to electron-hole plasma, rather than excitons. Hence, the corresponding PL intensity $I_P \propto f_e(E_e)f_h(E_h)$, where $f_e$ and $f_h$ are the electron and hole distribution functions obeying Fermi-Dirac statistics (with respective quasi-Fermi levels), and the corresponding electron ($E_e$) and hole ($E_h$) energies are measured from the conduction and valence band edges, respectively. As energy of the photo-excited carriers are much higher in this regime than the respective quasi-Fermi levels, we obtain (*27*)

$I_P \propto e^{[-(E-E_{g0})/K_BT]}$,

Hence,

$$\log(I_P) = \text{constant} - \frac{E - E_{g0}}{K_BT} \qquad (2)$$

On the other hand, if the hot luminescence is contributed from excitons, the intensity ($I_P$) is proportional to the number ($n$) of excitons present at that energy. Since $n$ follows Bose-Einstein distribution, for $E \gg K_BT$, we obtain:

$$\log(I_P) = \text{constant} - \frac{E}{K_BT} \qquad (3)$$

Equations (2) and (3) indicate that irrespective of the origin of the luminescence being associated with e-h plasma or excitons, the hot luminescence tail associated with the PL peaks are expected to decay exponentially with energy, and the corresponding carrier temperature can be extracted from the slope of a plot of $I_P$ (in log scale) and $E$.

Fig. 4 shows that the extracted carrier temperature from the tail of 2.29 eV peak is 510 ($\pm$10) K for an incident laser power of 0.26 mW. This corresponds to the carrier temperature before relaxation through optical phonons. The only possible relaxations in this regime are through carrier-carrier and carrier-acoustic phonon scattering (supporting information S4). The temperature extracted from the tail of the $A_{1s}$ peak (at 1.87 eV) is 363K$\pm$3K. Reducing the power level to 0.026mW reduces the extracted temperature of the $A_{1s}$ exciton to around 290K (inset of Fig. 4), which is the laboratory ambient temperature.

There are two independent ways in which the above extraction of carrier temperature can be ambiguous. First, proximity of other peaks may flatten the hot luminescence tail, and the extracted temperature in that case will be overestimated. In these situations, proper deconvolution of peaks is necessary before temperature extraction (supporting information S5). Second, we have assumed that the broadening of the photoluminescence peak results in completely by the Boltzmann-type tail, although in reality, there are other inhomogeneous contributions from excitation-induced and phonon-induced broadening (*28*). However, we show in Supporting Information S6, that at room temperature and above, the hot luminescence tail is the dominating broadening mechanism, and hence the temperature extraction is reliable to a first order. However, for accurate temperature extraction at low temperatures, one should subtract appropriate correction factors arising from other broadening mechanisms (Supporting Information S6).

The photo-carriers relax from the excited coherent state to the incoherent 1s state through emission of optical phonons, as schematically shown in Fig. 5. The optical phonons cannot transport the received heat efficiently due to poor group velocity. Instead, they decay through the acoustic phonons giving rise to anharmonic effects including a broadening ($\Gamma$) and a red shift ($\rho$) of the Raman peak (*29*):

$$I_R \propto \frac{\Gamma(T)}{[\omega_0 + \rho(T) - \omega]^2 + \Gamma^2(T)} \times [n(\omega, T) + 1] \qquad (4)$$

where $\omega_0$ is the peak without any anharmonic effect and $n(\omega, T)$ is the Bose-Einstein occupation number. In Fig. 6(a), we plot the Raman intensities at different incident laser powers, indicating red shift and broadening of both $A_{1g}$ and $E^1_{2g}$ peaks. We observe [Fig. 6(b)-(c)] that the $A_{1g}$ peak shows stronger broadening and red shift compared with the $E^1_{2g}$ peak. This is in agreement with the fact that the out of plane $A_{1g}$ phonon is the primary contributor to electron-phonon scattering (*30*), particularly at lower laser powers. However, at higher incident power, the $E^1_{2g}$ phonon branch also opens up a channel for energy relaxation.

Now, there are many different choices of phonon combinations through which an optical phonon can decay, as long as the energy and momentum are conserved. We choose only those phonon combinations which possess relatively large density of states (*31*), and use a modified Klemens model (*32*) for modeling the peak broadening. For example, FWHM of the $A_{1g}$ peak is obtained as (Supporting Information S7):

$$\Gamma(T) = \Gamma_0 + 2\alpha n^{LA(M)}(244, \sigma_1 T) + 2\alpha n^{TA(M)}(160, \sigma_2 T) + 2\beta n^{LA}(202, \sigma_3 T)$$
$$+ 2\gamma n^{E'}(384, \sigma_4 T) + 2\gamma n^{LA,TA}(20, \sigma_5 T) \qquad (5)$$

where $\Gamma_0$ is the broadening due to background, and the superscripts in $n$ indicate the respective phonon branches. $\sigma_i T$ (with $\sigma_i < 1$) is the temperature of the i$^{th}$ phonon branch where the energy is transferred to. The red shift of the Raman peak follows a linear relationship with a change in the temperature: $\rho(T) = \xi \Delta T$. Taking the peak position at a laser power of 0.026 mW to correspond to 290 K (from inset of Fig. 4), and using the value of $\xi$ ($= -1.30 \times 10^{-2}$cm$^{-1}$K$^{-1}$) for the A$_{1g}$ peak from recent literature (*33*), we first estimate the temperature of the A$_{1g}$ phonon branch. Next, with the estimated temperature and Eq. (5), we find that we can self-consistently fit the FWHM data at the corresponding laser powers [Fig. 6(d)]. The best fit is obtained with $\Gamma_0 = 0.96, \alpha = 0.94, \beta = 0.04, \gamma = 0.02$. This shows that the primary relaxation path are the phonon branches at the M point. Also, we obtained $\sigma_1 = \sigma_2 = 1$ implying that the temperature of the A$_{1g}$ phonon is almost equal to the decaying acoustic branches, which is in agreement with the fact that these acoustic branches at the M point are almost flat possessing very small group velocity (*31*) and thus cannot carry away the heat efficiently, rather in turn transfer the heat to other acoustic branches. The extracted A$_{1g}$ and E$^1_{2g}$ temperatures at different laser powers are plotted in Fig. 6(e). The different mechanisms discussed above are summarized in Fig. 5.

In conclusion, using off-resonant excitation, we demonstrated linear polarization resolved hot photoluminescence at room temperature in monolayer MoS$_2$, MoSe$_2$, WSe$_2$ and WS$_2$ reflecting valley coherence close to excitation energy due to lack of sufficient optical phonon scattering. We have also shown strong room temperature valley coherence in monolayer MoSe$_2$ by resonantly exciting the 1s peak at energies close to excitation. Generation of valley coherence is important from the point of view of valleytronics, where manipulation of valley degree of freedom by external stimulation is required. Hence demonstration of such coherence at room temperature, on one hand, is technologically important due to increased hopes for room temperature valleytronics, and on the other hand, makes laboratory experiments more flexible in terms of laser wavelength and sample temperature requirements. We also found the lower bound of the A exciton binding energy in monolayer MoS$_2$ has been extracted to be ~0.42 ($\pm$0.02) eV. Using a combination of hot luminescence tail and anharmonic effects in Raman peaks, the extracted steady state temperature of the carriers before optical phonon relaxation, after relaxation to 1s level and that of the emitted A$_{1g}$ optical phonons are 510 ($\pm$10) K, 363 ($\pm$3) K, and 337 ($\pm$15) K respectively, for an incident laser power of 0.26 mW on monolayer MoS$_2$. Such in-situ and non-destructive extraction of temperature of the carriers as well as the different phonon branches is important as it is generally non-trivial to directly find out the temperature of the electron (or hole) and phonon systems separately in the semiconductor under non-equilibrium. The technique can also be easily extended to other systems as well, for example in devices under high electric bias.

## Methods

**Sample preparation and measurement setup.** Monolayer TMDs are exfoliated from bulk crystals (procured from *2D Semiconductors*) on cleaned Si wafers covered by 285 nm SiO$_2$. The detailed characterization of the different monolayer materials (optical image, thickness measurement by AFM, and Raman characterization) are provided in Supporting Information S1. The experimental setup is shown in 1(a), where a vertically polarized 532 nm or 785nm laser light is passed through a half-wave plate kept at angle θ. This generates either vertically ($\theta = 0°$) or horizontally ($\theta = 45°$) polarized light that falls on the sample through a 100X objective. The back scattered light is passed through a vertical analyzer, and finally collected in a Peltier cooled CCD detector. All measurements are performed at 290 K.

**Acknowledgement**

The authors acknowledge the support of nano-fabrication and characterization facilities at Center for Nano Science and Engineering. KM would like to acknowledge support of the start-up grant from Indian Institute of Science, and the research grant under Ramanujan fellowship, Department of Science and Technology, Government of India.

**Supporting Information Available:** Supporting information available on material characterization, symmetry analysis with $MoS_2$ Raman peaks, literature summary on exciton binding energy, analysis of energy relaxation close to excitation, gate voltage dependent PL peak shift, inhomogeneous PL peak broadening mechanisms, and anharmonic decay model of phonons.

**Corresponding author**

*Email: kausikm@ece.iisc.ernet.in

# References

1. Mak, K. F.; Lee, C.; Hone, J.; Shan, J.; Heinz, T. F. Atomically Thin MoS2: A New Direct-Gap Semiconductor. *Phys. Rev. Lett.* **2010,** *105,* 136805.

2. Splendiani, A.; Sun, L.; Zhang, Y.; Li, T.; Kim, J.; Chim, C.; Galli, G.; Wang, F. Emerging photoluminescence in monolayer MoS2. *Nano Lett.* **2010,** *10,* 1271.

3. Xia, F.; Wang, H.; Xiao, D.; Dubey, M.; Ramasubramaniam, A. Two-dimensional material nanophotonics. *Nature Photon.* **2014,** *8,* 899-907.

4. Baugher, B.; Churchill, H.; Yang, Y.; Jarillo-Herrero, P. Optoelectronic devices based on electrically tunable p-n diodes in a monolayer dichalcogenide. *Nat. Nanotechnol.* **2014,** *9,* 262-267.

5. Pospischil, A.; Furchi, M.; Mueller, T. Solar-energy conversion and light emission in an atomic monolayer p – n diode. *Nat. Nanotechnol.* **2014,** *9,* 257-261.

6. Ross, J.; Klement, P.; Jones, A.; Ghimire, N.; Yan, J.; Mandrus, D. G.; Taniguchi, T.; Watanabe, K.; Kitamura, K.; Yao, W.; Cobden, D. H.; Xu, X. Electrically tunable excitonic light-emitting diodes based on monolayer WSe2 p-n junctions. *Nat. Nanotechnol.* **2014,** *9,* 268-272.

7. Lopez-Sanchez, O.; Lembke, D.; Kayci, M.; Radenovic, A.; Kis, A. Ultrasensitive photodetectors based on monolayer MoS2. *Nat. Nanotechnol.* **2013,** *8,* 497-501.

8. Cao, T.; Wang, G.; Han, W.; Ye, H.; Zhu, C.; Shi, J.; Niu, Q.; Tan, P.; Wang, E.; Liu, B.; Feng, J. Valley-selective circular dichroism of monolayer molybdenum disulphide. *Nat. Commun.* **2012,** *3,* 887.

9. Xiao, D.; Liu, G. B.; Feng, W.; Xu, X.; Yao, W. Coupled spin and valley physics in monolayers of MoS 2 and other group-VI dichalcogenides. *Phys. Rev. Lett.* **2012,** *108,* 196802.

10. Mak, K. F.; Mcgill, K. L.; Park, J.; Mceuen, P. L. The valley Hall effect in MoS 2 transistors. *Science* **2014,** *344,* 1489-1492.

11. Aivazian, G.; Gong, Z.; Jones, a. M.; Chu, R. L.; Yan, J.; Mandrus, D. G.; Zhang, C.; Cobden, D.; Yao, W.; Xu, X. D. Magnetic Control of Valley Pseudospin in Monolayer WSe2. *Nat. Phys.* **2015,** *11,* 148-152.

12. Srivastava, A.; Sidler, M.; Allain, A. V.; Lembke, D. S.; Kis, A.; Imamoğlu, A. Valley Zeeman effect in elementary optical excitations of monolayer WSe 2. *Nat. Phys.* **2014,** *11,* 147.

13. Zhang, Y. J.; Oka, T.; Suzuki, R.; Ye, J. T.; Iwasa, Y. Electrically switchable chiral light-emitting transistor. *Science* **2014,** *344,* 725-728.

14. Ramasubramaniam, A. Large excitonic effects in monolayers of molybdenum and tungsten dichalcogenides. *Phys. Rev. B* **2012,** *86,* 115409.

15. Chernikov, A.; Berkelbach, T. C.; Hill, H. M.; Rigosi, A.; Li, Y.; Aslan, O. B.; Reichman, D. R. H. M. S. H. T. F. Exciton Binding Energy and Nonhydrogenic Rydberg Series in Monolayer WS2. *Phys. Rev. Lett.* **2014,** *113,* 076802.

16. He, K.; Kumar, N.; Zhao, L.; Wang, Z.; Mak, K. F.; Zhao, H.; Shan, J. Tightly Bound Excitons in Monolayer WSe2. *Phys. Rev.Lett.* **2014,** *113,* 026803.

17. Hill, H.; Rigosi, A.; Roquelet, C.; Chernikov, A.; Berkelbach, T.; Reichman, D.; Hybertsen, M.; Brus, L.; Heinz, T. Observation of excitonic Rydberg states in monolayer MoS2 and WS2 by photoluminescence excitation spectroscopy. *Nano Lett.* **2015,** *15,* 2992-2997.

18. Ye, Z.; Cao, T.; Brien, K. O.; Zhu, H.; Yin, X.; Wang, Y.; Louie, S. G.; Zhang, X. Probing excitonic dark states in single-layer tungsten disulphide. *Nature* **2014,** *513,* 214-218.

19. Ugeda, M. M.; Bradley, A. J.; Shi, S.-F.; da Jornada, F. H.; Zhang, Y.; Qiu, D. Y.; Mo, S.-K.; Hussain, Z.; Shen, Z.-X.; Wang, F.; Louie, S. G.; Crommie, M. F. Observation of giant bandgap renormalization and excitonic effects in a monolayer transition metal dichalcogenide semiconductor. *Nat. Mater.* **2014,** *13,* 1091-1095.

20. Jones, A. M.; Yu, H.; Ghimire, N. J.; Wu, S.; Aivazian, G.; Ross, J. S.; Zhao, B.; Yan, J.; Mandrus, D. G.; Xiao, D.; Yao, W.; Xu, X. Optical generation of excitonic valley coherence in monolayer WSe2. *Nat. Nanotechnol.* **2013,** *8,* 634-638.

21. Mai, C.; Barrette, A.; Yu, Y.; Semenov, Y. G.; Kim, K. W.; Cao, L.; Gundogdu, K. Many-body effects in valleytronics: Direct measurement of valley lifetimes in single-layer MoS2. *Nano Lett.* **2014,** *14,* 202-206.

22. Kozawa, D.; Kumar, R.; Carvalho, A.; Amara, K. K.; Zhao, W.; Wang, S.; Toh, M.; Ribeiro, R. M.; Neto, A. H. C.; Matsuda, K.; Eda, G. Photocarrier relaxation pathway in two-dimensional semiconducting transitino metal dichalcogenides. *Nat. Commun.* **2014,** *5,* 1-7.

23. Wang, G.; Bouet, L.; Lagarde, D.; Vidal, M.; Balocchi, A.; Amand, T.; Marie, X.; Urbaszek, B. Valley dynamics probed through charged and neutral exciton emission in monolayer WSe2. *Phys. Rev. B* **2014,** *90,* 075413.

24. Koch, S.; Kira, M.; Khitrova, G.; Gibbs, H. M. Semiconductor excitons in new light. *Nat. Mater.* **2006,** *5,* 523-531.

25. Zemskii, V.; B, Z.; D, M. Polarization of hot photoluminescence in semiconductors of GaAs type. *J. Exp. Theor. Phys.* **1976,** *24,* 96-99.

26. Klingshirn, C. F. *Semiconductor Optics;* Springer, 2012.

27. Lyon, S. Spectroscopy of hot carriers in semiconductors. *JOL* **1986,** *35,* 121-154.

28. Moody, G.; Kavir Dass, C.; Hao, K.; Chen, C.-H.; Li, L.-J.; Singh, A.; Tran, K.; Clark, G.; Xu, X.; Berghäuser, G.; Malic, E.; Knorr, A.; Li, X. Intrinsic homogeneous linewidth and broadening mechanisms of excitons in monolayer transition metal dichalcogenides. *Nat. Commun.* **2015,** 8315.

29. Menéndez, J.; Cardona, M. Temperature dependence of the first-order Raman scattering by phonons in Si, Ge, and -Sn: Anharmonic effects. *Phys. Rev. B* **1984,** *29*.

30. Kaasbjerg, K.; Thygesen, K. S.; Jacobsen, K. Phonon-limited mobility in n-type single-layer MoS 2 from first principles. *Phys. Rev. B* **2012,** *85*.

31. Sanchez, M.; Wirtz, L. Phonons in single-layer and few-layer MoS2and WS2. *Phys. Rev. B* **2011,** *84,* 155413.

32. Klemens, P. G. Anharmonic decay of optical phonons. *Phys. Rev. B* **1966,** *148*.

33. Yan, R.; Simpson, J. R.; Simone, B.; Brivio, J.; Watson, M.; Wu, X.; Kis, A.; Luo, T.; Hight Walker, A. R.; Xing, H. G. Thermal Conductivity of Monolayer Molybdenum Disulfide Obtained from Temperature-Dependent Raman Spectroscopy. *ACS Nano* **2014,** *8,* 986-993.

**Figure captions:**

**Figure 1.** Polarization resolved PL and Raman signal in monolayer $MoS_2$. (a) Schematic of the setup for linear polarization resolved PL and Raman measurement. (b) Linear polarization resolved PL signal showing near zero degree of polarization at the A and B peak, however close to excitation energy, co-polarization (VV) setup shows more intensity than cross polarization (HV). The sharp peaks close to excitation energy are backscattered Raman signal from $MoS_2$ and Si. (c) Out of plane $A_{1g}$ Raman peak is strongly linearly polarized, while the in-plane $E^1_{2g}$ peak is not, owing to symmetry of the vibrations (Supporting Information 1).

**Figure 2.** Room temperature valley coherence and polarization resolved hot luminescence in $MoS_2$, $WSe_2$ and $WS_2$. (a)-(c) Linear polarization resolved hot luminescence signal indicating valley coherence. The dashed lines are guide to eyes for background PL signals resolved from Raman peaks. (d) Extracted degree of linear polarization $\rho = (I_{VV} - I_{HV})/(I_{VV} + I_{HV})$ from a-c. (e) Intrinsic degree of polarization in monolayer TMDs as a function of $\boldsymbol{k}$ measured from $\boldsymbol{K(K')}$ (perfect polarization obtained only at $\boldsymbol{k} = 0$). The dashed arrows indicate the excitation position in the Brillouin zone for $A_{1s}$ states of $MoS_2$, $WSe_2$ and $WS_2$. (f)-(h), Schematic diagram showing a monochromatic laser excites multiple excitonic levels in parallel, with lower excitonic level gets excited farther from $\boldsymbol{K(K')}$ point. Considering A exciton, 2.33 eV laser excites just above continuum level in $MoS_2$, well above continuum in $WSe_2$, and below continuum limit in $WS_2$.

**Figure 3.** Room temperature valley-coherence in $MoSe_2$ with resonant and off-resonant excitation. (a) Schematic of off-resonant excitation in monolayer $MoSe_2$. (b) Unpolarized 1s peak PL spectrum as a consequence of 2.33 eV (532 nm) excitation. (c) Linearly polarized PL spectrum

close to excitation. (d) Schematic of resonant excitation in monolayer MoSe$_2$. (e) Strong linear polarization of 1s peak as a result of resonant excitation by 1.58 eV (785 nm) laser.

**Figure 4.** Carrier temperature extraction from hot luminescence tail in monolayer MoS$_2$. The temperatures extracted from the exponential hot luminescence tail are (i) 510 ($\pm$10) K for hot carriers before optical phonon relaxation and (ii) 363 ($\pm$3) K after relaxation to A$_{1s}$ state, for a 0.26 mW incident power. Inset, Extracted carrier temperature at A$_{1s}$ state is 290 K at 0.026 mW power.

**Figure 5.** Summary of valley coherent hot carrier generation and relaxation mechanisms. Schematic representation of coherent valley excitation by linearly polarized light (in green), coherent recombination before optical phonon scattering and thus polarization resolved hot PL generation (in green), subsequent relaxation to 1s states through optical phonon emission (in blue), and valley incoherent (hence, unpolarized) strong PL at 1s state (in red).

**Figure 6.** Phonon temperature extraction from anharmonic effects in Raman peaks in monolayer MoS$_2$. (a) Raman spectra showing A$_{1g}$ and E$^1_{2g}$ peaks at different incident power levels. (b)-(c), Peak red shift and FWHM broadening of Raman peaks at different incident power levels. Error bars calculated are from laser spots on multiple positions of different flakes. (d) Self-consistent fitting of peak shift and broadening with phonon temperatures. The red lines are obtained from the anharmonic broadening model. (e) Extracted phonon temperatures from (d) as a function of incident laser power.

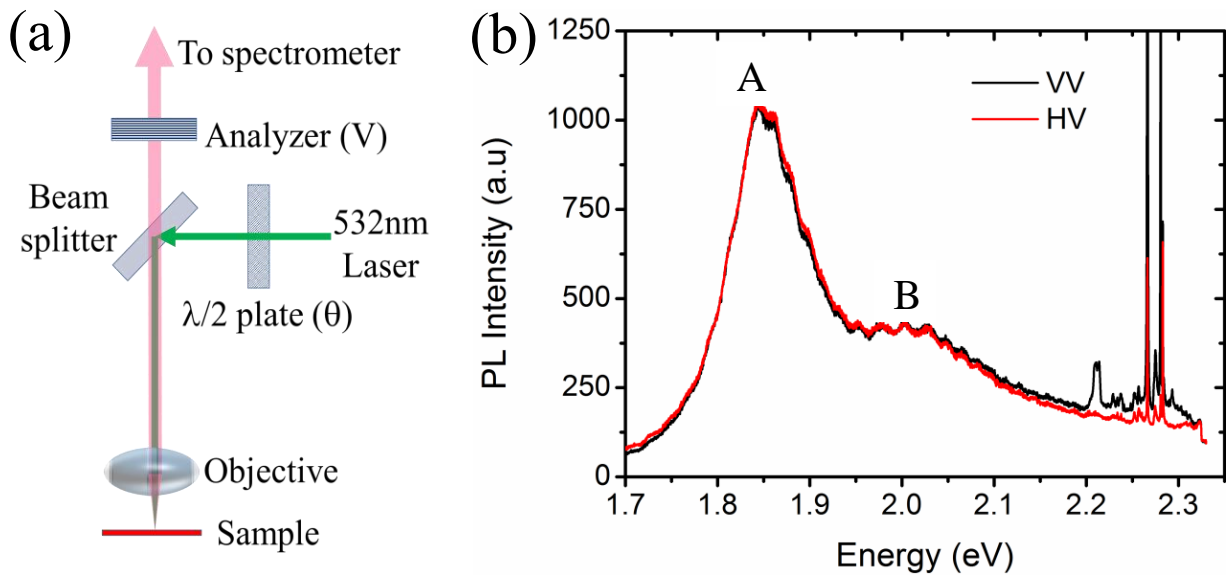

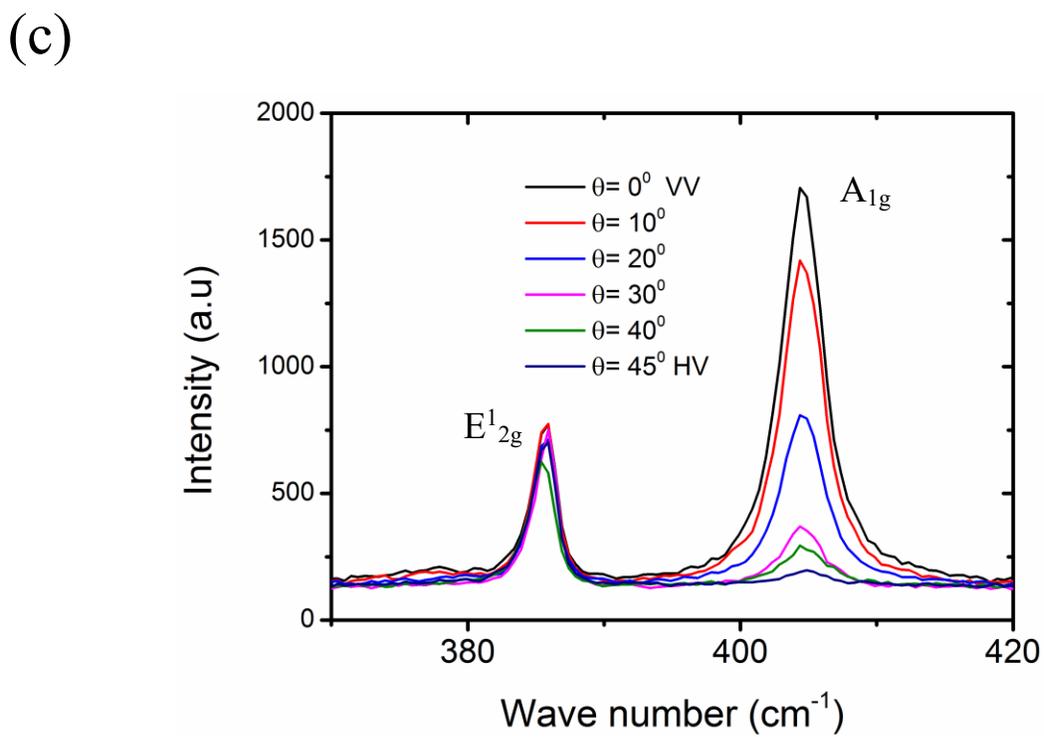

Figure 1

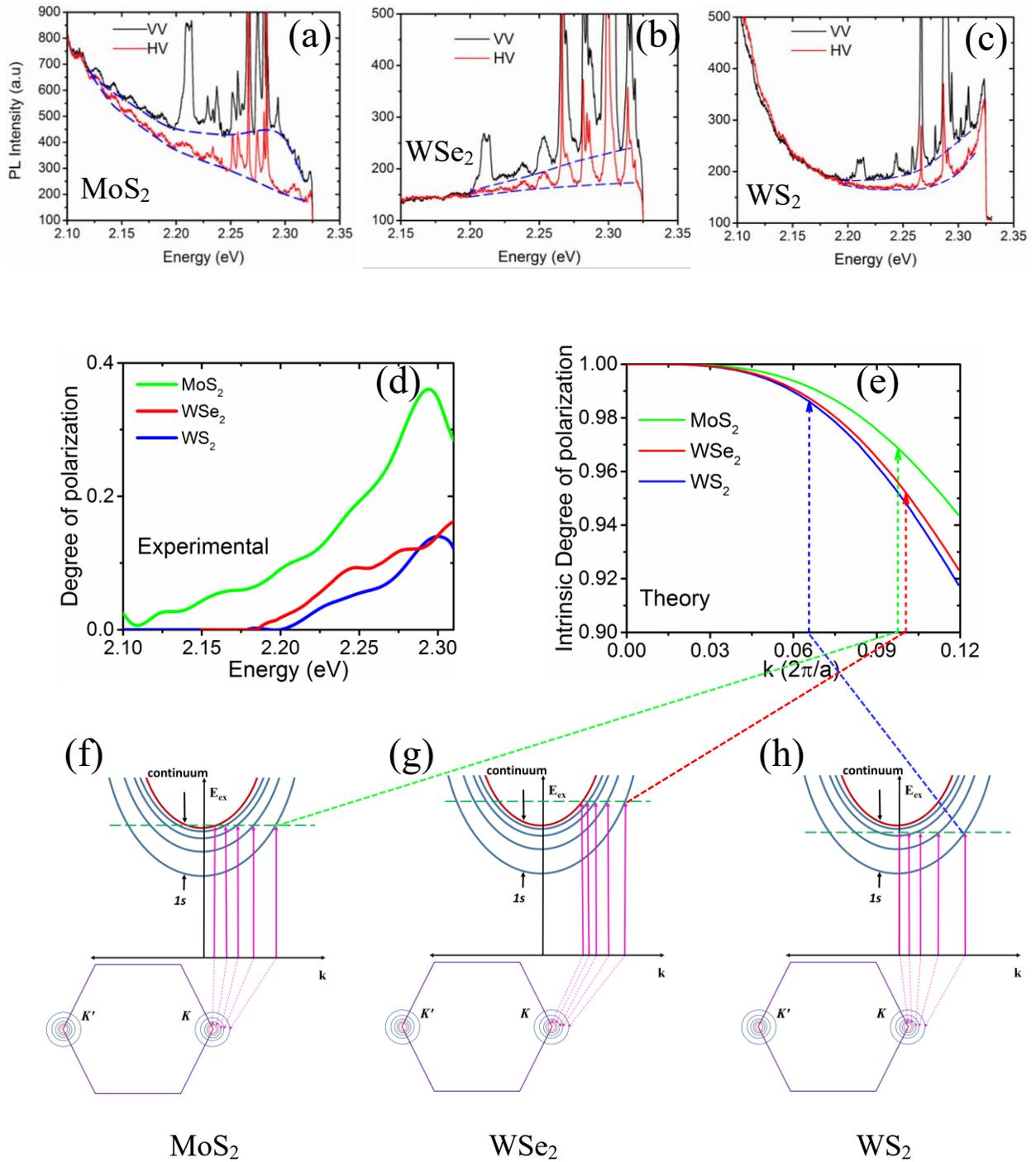

Figure 2

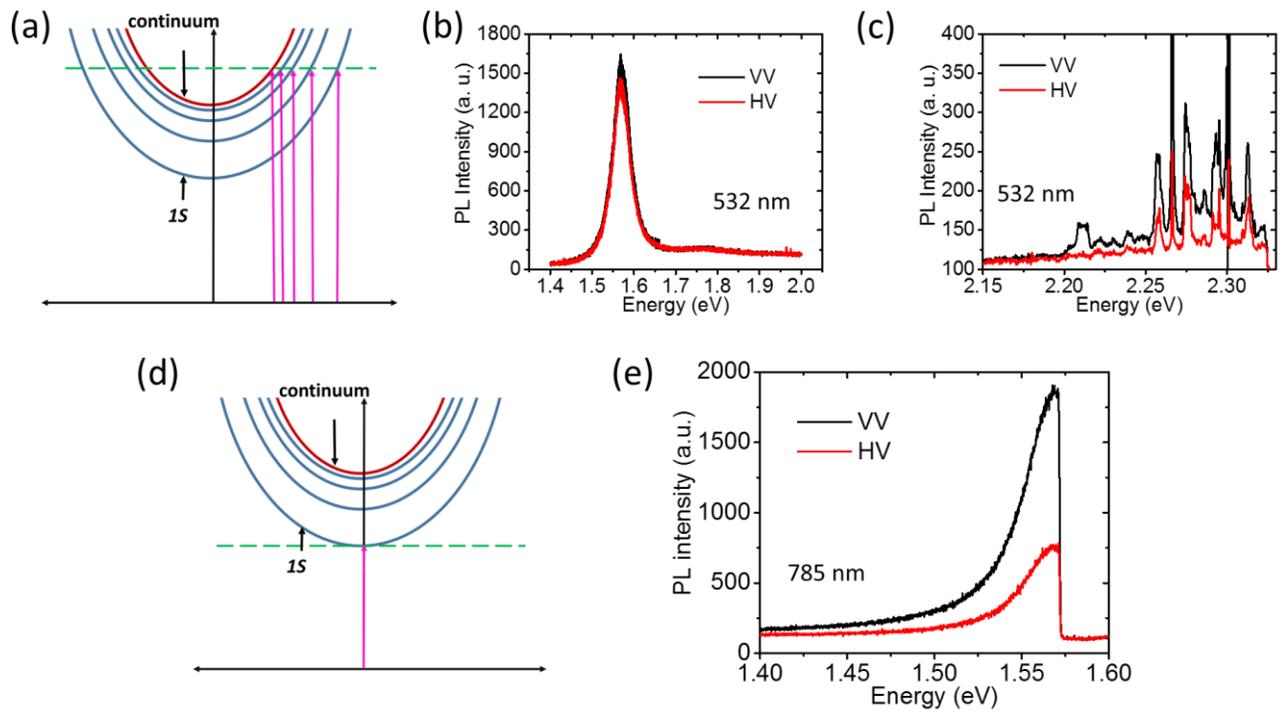

Figure 3

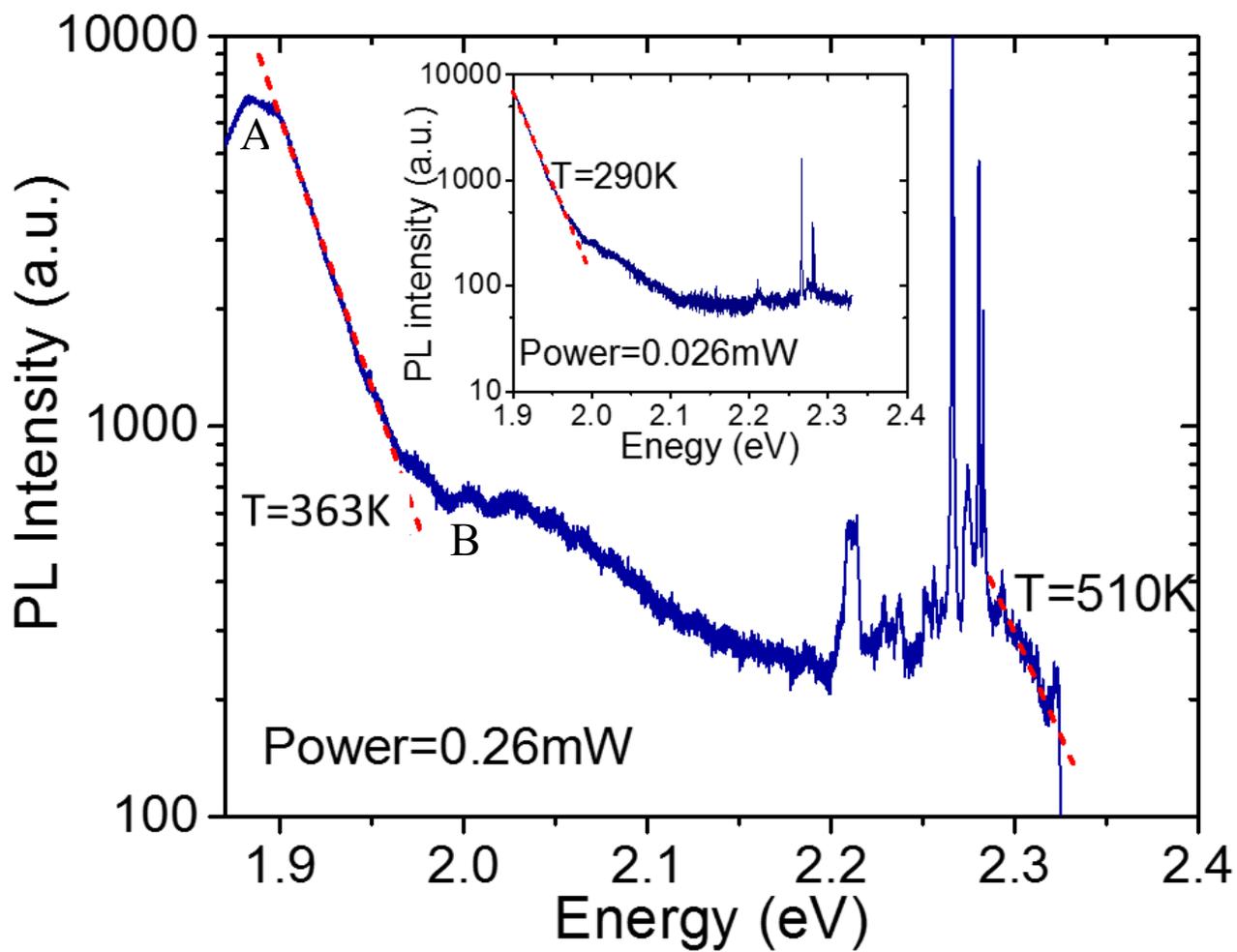

Figure 4

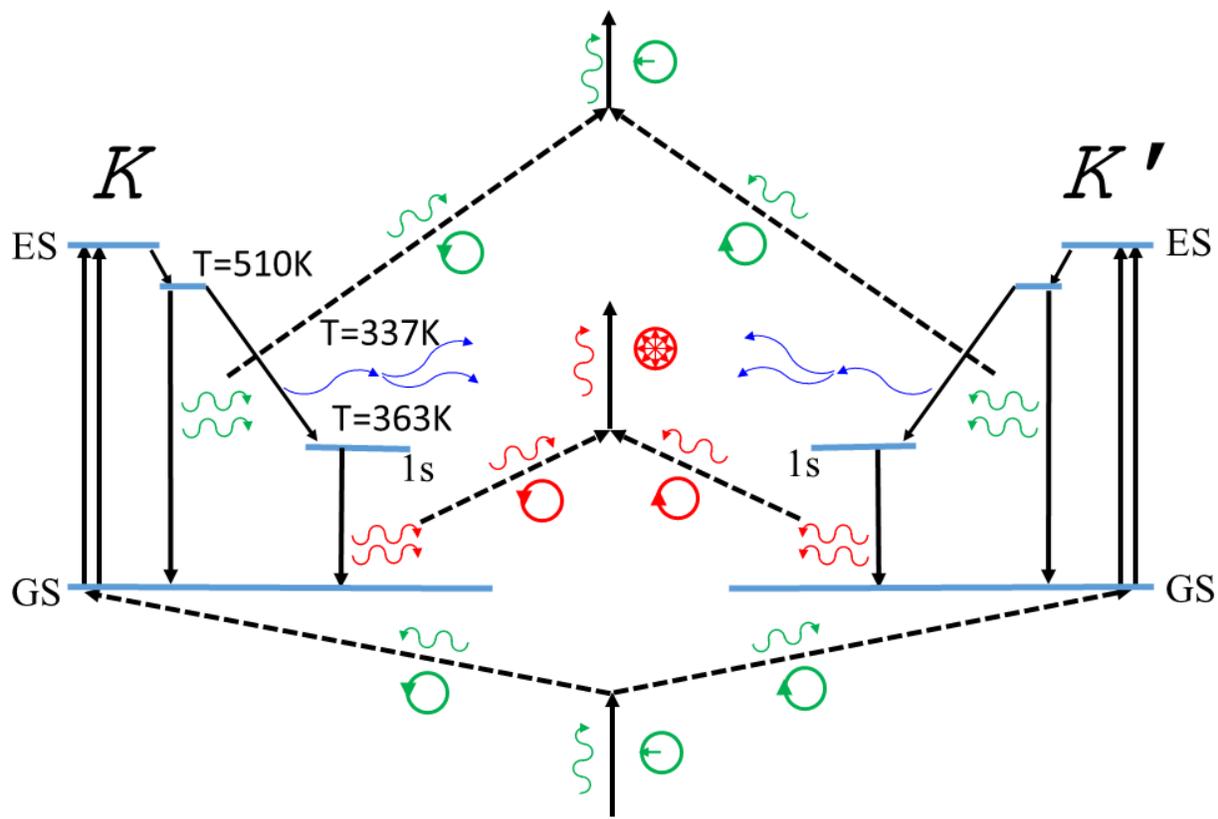

Figure 5

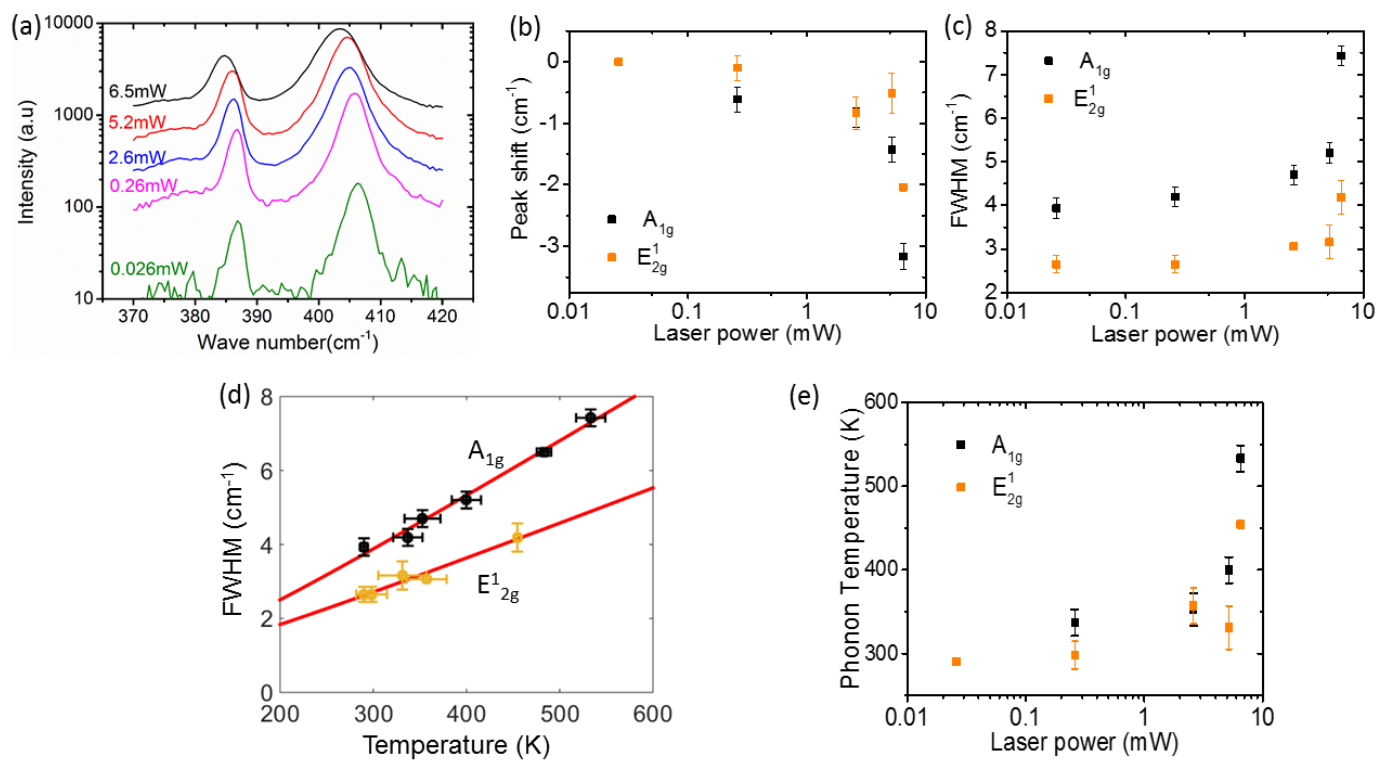

Figure 6

# Supporting information for

# Valley coherent hot carriers and thermal relaxation in monolayer transition metal dichalcogenides

*Sangeeth Kallatt[1,2,3], Govindarao Umesh[3], and Kausik Majumdar[1] \**

[1]Department of Electrical Communication Engineering, Indian Institute of Science, Bangalore 560012, India

[2]Center for NanoScience and Engineering, Indian Institute of Science, Bangalore 560012, India

[3]Department of Physics, National Institute of Technology Karnataka, Mangalore 575025, India

## S1. Material characterization of monolayer MoS$_2$, MoSe$_2$, WS$_2$, WSe$_2$

The figure below provides material characterization (optical contrast, AFM, and Raman spectroscopy) of different monolayer materials, with the columns representing MoS$_2$, MoSe$_2$, WS$_2$ and WSe$_2$, respectively from left to right.

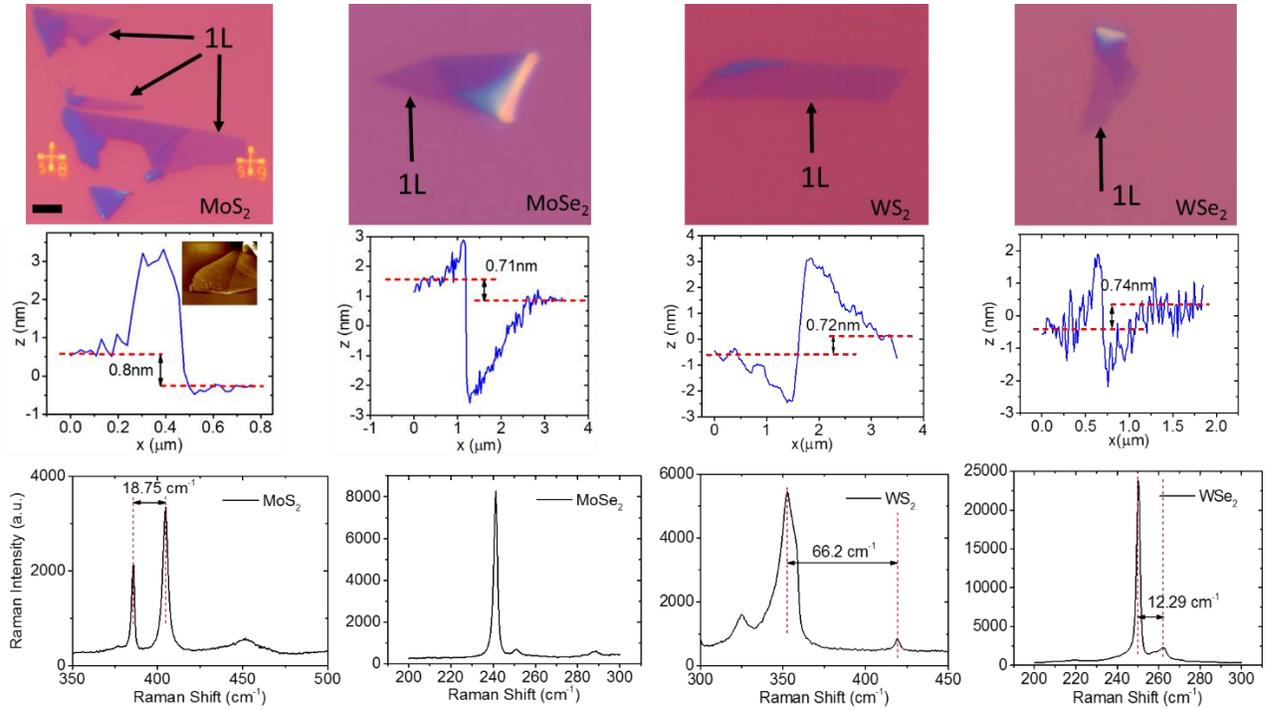

Figure S1. Row 1: Optical image of flakes showing optical contrast of monolayer flakes. Row 2: Thickness of monolayer flakes using AFM. Row 3: Raman spectroscopy of monolayer flakes.

## S2. Polarization resolved Raman peaks in MoS₂ – symmetry analysis

The incident vertically polarized light is passed through a half wave plate kept at an angle $\theta/2$ (and hence rotates the polarization by angle $\theta$), which then falls on the sample and the Raman scattered signal is passed through a vertical analyzer (as shown in Fig. 1(a) in the main text). From symmetry, the $A_{1g}$ and $E^1_{2g}$ peak intensities are given by (using the corresponding Raman tensors)

$$I_{A_{1g}} \propto \left| [0\ 1\ 0] \begin{bmatrix} a & 0 & 0 \\ 0 & a & 0 \\ 0 & 0 & b \end{bmatrix} \begin{bmatrix} \sin\theta \\ \cos\theta \\ 0 \end{bmatrix} \right|^2 = a^2 \cos^2\theta \qquad (s1)$$

and

$$I_{E^1_{2g}} \propto \left| [0\ 1\ 0] \begin{bmatrix} 0 & d & 0 \\ d & 0 & 0 \\ 0 & 0 & 0 \end{bmatrix} \begin{bmatrix} \sin\theta \\ \cos\theta \\ 0 \end{bmatrix} \right|^2 + \left| [0\ 1\ 0] \begin{bmatrix} d & 0 & 0 \\ 0 & -d & 0 \\ 0 & 0 & 0 \end{bmatrix} \begin{bmatrix} \sin\theta \\ \cos\theta \\ 0 \end{bmatrix} \right|^2 = d^2 \qquad (s2)$$

Clearly, the $A_{1g}$ peak is strongly polarized, with maximum intensity when $\theta = 0°$ (co-polarization or VV setup) and zero intensity when $\theta = 90°$ (cross-polarization or HV setup). On the other hand, the $E^1_{2g}$ peak is insensitive to the polarization angle. In (s2), the intensities are added together, indicating an incoherence between the two components (which supports the experimental data).

## S3. Reported binding energy of excitons in monolayer TMDs in recent literature

| Sl. No. | Material | A or B exciton | Binding energy (eV) | Method | Ref. |
|---|---|---|---|---|---|
| 1 | $MoS_2$ | B | 0.44 | PL | [1] |
| 2 | $MoS_2$ | A | 0.3 | STM | [2] |
| 3 | $MoS_2$ | A | 0.57 | Photocurrent | [3] |
| 4 | $WS_2$ | A | 0.32 | Reflectance | [4] |
| 5 | $WS_2$ | A | 0.32 | PL | [1] |
| 6 | $WS_2$ | A | 0.7 | Two photon excitation (dark states) | [5] |
| 7 | $WS_2$ | A | 0.79 | Reflectance/absorption | [6] |
| 8 | $WSe_2$ | A | 0.37 | Linear absorption and two photon | [7] |
| 9 | $WSe_2$ | A | 0.4 | STM | [2] |
| 10 | $WSe_2$ | A | 0.47 | SHG | [8] |
| 11 | $MoSe_2$ | A | 0.55 | STM | [9] |

## S4. Nature of depolarization close to excitation energy

The depolarization mechanism close to the laser excitation can be understood by analyzing depolarization dynamics from each valley. One can build a simple phenomenological model including intra-valley and inter-valley scattering as follows:

$$\frac{dP_K}{dt} = -\alpha_1 P_K - \alpha_2(P_K - P_{K'}) \qquad (1)$$

$$\frac{dP_{K'}}{dt} = -\alpha_1 P_{K'} - \alpha_2(P_{K'} - P_K) \qquad (2)$$

where $P_K$ and $P_{K'}$ are the valley dependent polarizations, $\alpha_1$ and $\alpha_2$ are intra-valley and inter-valley scattering parameters. Inter-valley scattering is dominated by optical phonons and acoustic phonons with large phonons. Thus, at energy values very close to the excitation, inter-valley scattering can be neglected. This reduces a simple exponential decay of the polarization as:

$$P_K(t) = P_K(0)e^{-\alpha_1 t} \qquad (3)$$

where $P_K(0)$ is the initial polarization at excitation energy. The intra-valley scattering parameter $\alpha_1$ includes fast carrier-carrier scattering and also scattering due to close to zone center acoustic phonons. The above analysis indicates that close to excitation, polarization decays exponentially with time. Now, for steady state measurements with scanning energies, the detector collects photons at different energies, which are emitted by gradually relaxing excitons at different time scales. Hence, the time scale in some sense mapped to the energy scale, explaining depolarization with energy from the excitation energy.

## S5. Accurate extraction of carrier temperature – gate voltage dependent PL peak shift

One should be careful in applying the technique of temperature extraction from the luminescence tail, when multiple peaks are in close proximity. In this case, the slope can be flattened due to contribution from other peaks, resulting in predicting higher temperature. Such a situation occurs when the $MoS_2$ sample is doped which results in a trion peak and an exciton peak with a separation of ~30 meV. Fig. S2 shows such an example where the sample was initially n-type doped. The PL peak around 1.845 eV has a primary contribution from the trion peak, with a small contribution from the exciton, which flattens the slope significantly. When we apply a negative bias from the back gate, the effective doping in the sample is reduced, and we are able to suppress the trion peak, blue shifting the PL peak by around 30 meV, and now the slope has primary contribution from the exciton – hence the extracted temperature will be more accurate.

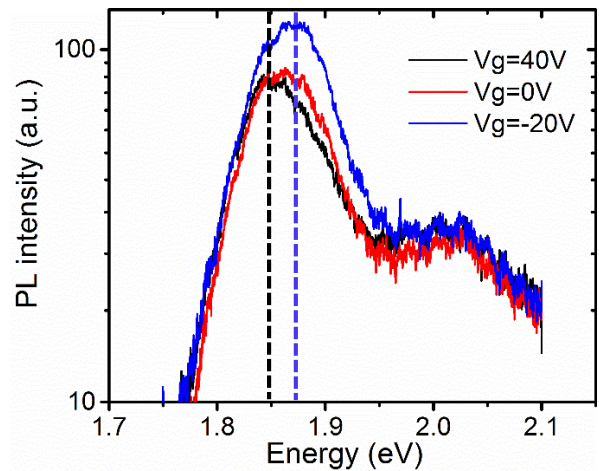

## S6. Inhomogeneous PL peak broadening mechanisms

The total broadening of the PL peak can be given as:

$$\gamma(T, N_x) = \gamma_0(0) + \gamma_x N_x + \gamma_0(T)$$

where the temperature independent term arises from excitation induced broadening [10], and the temperature dependent term (the last term) can be written as a sum of exciton-phonon scattering effects and the exponential tail of the hot luminescence as:

$$\gamma_0(T) = \gamma' T + \log_e(2)\, k_B T$$

Based on the data in ref [10], $\gamma'$ is obtained to be 60 meV/K. Interestingly, the second term by itself produces a slope of $\log_e(2)\, k_B$, which is also exactly equal to 60 meV/K. The excellent fit of the calculated hot luminescence induced broadening with the experimental data in ref. [10] is shown below.

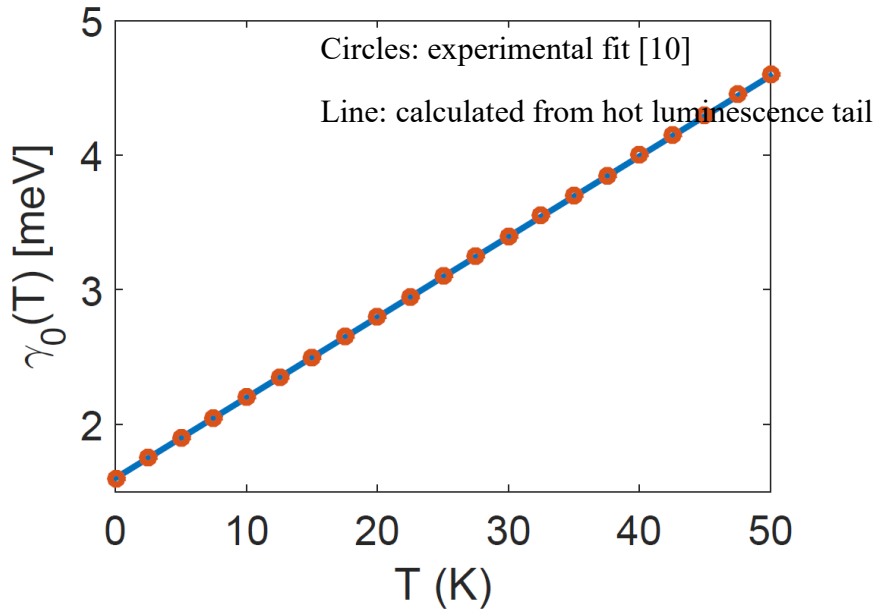
30

Note that, the broadening due to the tail of the luminescence is intrinsic and is independent of any fitting. Thus, we can conclude that at the measurement temperature of ref. [10] (T=10K), the temperature dependent broadening can be completely attributed to the exponential tail, and any broadening due to exciton-phonon interaction can be safely neglected.

Assuming broadening due to exponential tail is much larger than due to exciton-phonon interaction even at higher temperature (as both are linear in temperature), and using the values of $\gamma_0$ and $\gamma_x$ as provided in ref. [10], the broadening at room temperature (T=300 K, $N_x$=$10^{11}$cm$^{-2}$) can be written as

$$\gamma(T = 300K, N_x = 10^{11} cm^{-2}) = 1.60 + 0.27 + \mathbf{18} \text{ (in meV)}$$

where the last term in bold is from the hot luminescence exponential tail and contributes to 90.6% to the total broadening. Higher the temperature, % contribution of the exponential tail increases (92.8% at 400K, and 94.1% at 500K).

Hence we conclude that at room temperature and above (which is the case for the present work), we can safely assume that, to a first order, the exponential tail of hot luminescence is the primary contributor to PL peak broadening. This is also evidenced by the excellent exponential fit to the PL tail over an order of magnitude of the PL intensity (as shown in main text). However, at very low temperature, the technique of temperature extraction would need to include correction factor.

## S7. Phonon decay and anharmonic broadening of Raman peak

The phonon-phonon interaction obeys conservation of momentum and energy. Now, the Raman peak corresponds to zone center ($q = 0$) phonon, so we expect zero sum of momentum of the phonons into which the parent phonon decays, and the sum of energy would be equal to the parent phonon. The broadening equation for A$_{1g}$ peak hence can be written with primary contribution from phonons with large density of states:

$$\Gamma(T) = \Gamma_0 + \alpha\{n^{LA(M)}(244, \sigma_1 T) + n^{TA(-M)}(160, \sigma_2 T)\}$$
$$+ \alpha\{n^{LA(-M)}(244, \sigma_1 T) + n^{TA(M)}(160, \sigma_2 T)\}$$
$$+ \beta\{n^{LA(+q_3)}(202, \sigma_3 T) + n^{LA(-q_3)}(202, \sigma_3 T)\} + \gamma\{n^{E'(+q_4)}(384, \sigma_4 T)$$
$$+ n^{LA,TA(-q_4)}(20, \sigma_5 T)\} + \gamma\{n^{E'(-q_4)}(384, \sigma_4 T) + n^{LA,TA(+q_4)}(20, \sigma_5 T)\}$$

where $q_3, q_4$ are unique values of momentum obtained from phonon dispersion relation. Noting that $E(+\boldsymbol{q}) = E(-\boldsymbol{q})$, we obtain equation (5) in the main text.

References:
[1] Hill et al, Nano Lett., 15, 2992 (2015).
[2] Chiu et al, Nature Comm., 6, 7666 (2015).
[3] Klots et al, Sci. Rep., 4, 6608 (2015).
[4] Chernikov et al, Phys. Rev. Lett., 113, 076802 (2014).
[5] Ye et al, Nature, 513, 214 (2014).
[6] Hanbicki et al, arXiv:1412.2156.
[7] He et al, Phys. Rev. Lett., 113, 026803 (2014).
[8] Wang et al, Phys. Rev. Lett., 114, 097403 (2015).
[9] Ugeda e al, Nature materials, 13, 1091 (2014).
[10] Moody et al, Nature Comm., 6, 8315 (2015).